\documentclass[aps,twocolumn,amssymb,amsmath]{revtex4}
\usepackage{graphicx}
\begin{document}

\title{2-GHz clock quantum key distribution over 260 km of standard telecom fiber}
\author{Shuang Wang,$^{1}$ Wei Chen,$^{1}$ Jun-Fu Guo,$^{2}$ Zhen-Qiang Yin,$^{1}$ Hong-Wei Li,$^{1}$
Zheng Zhou,$^{1}$ Guang-Can Guo,$^{1}$ and Zheng-Fu Han$^{1}$}

\address{
$^{1}$Key Laboratory of Quantum Information, University of
Science and Technology of China, CAS, Hefei 230026, China \\
$^2$Anhui Asky Quantum Technology Co.,Ltd., Wuhu 241002, China}

\begin{abstract}We report a demonstration of quantum key distribution (QKD) over a
standard telecom fiber exceeding 50 dB in loss and 250 km in length.
The differential phase shift QKD protocol was chosen and implemented
with 2 GHz system clock rate. By careful optimization of the 1-bit
delayed Faraday-Michelson interferometer and the use of the
super-conducting single photon detector (SSPD), we achieved a
quantum bit error rate below 2\% when the fiber length was no more
than 205 km, and of 3.45\% for the 260 km length fiber with 52.9 dB
loss. We also improved the quantum efficiency of SSPD to obtain high
key rate for 50 km length.\end{abstract}

\maketitle

\noindent Quantum key distribution (QKD) enables two remote
participants to share unconditionally secure keys based on the
principles of quantum physics \cite{bb84,gisin2002,scarani}.
Combined with one-time pad encryption, QKD is hopeful to effectively
end the cat and mouse game between the guardians of secrets and
their enemies \cite{broad}, and has become one of the most dynamic
research fields. After the past two decades of developments,
experimental QKD has achieved significant improvements, the
transmission distance from 32 cm \cite{first} to 250 km \cite{cow},
the speed (or system clock rate) from 200 Hz \cite{first} to 10 GHz
\cite{oe06,nphoton,townsend}.

With the dispersion-shifted fiber, Takesue et al. realized the first
QKD experiment over 42.1 dB channel loss and 200 km of distance
\cite{nphoton}. Then, with the ultra low loss fiber, Stucki et al.
implemented the first QKD experiment over 250 km of distance, but
the channel loss is still 42.6 dB \cite{cow}. In this letter,
focused on the transmission over the widely used standard (ITU-T
G.652) telecom fiber, of which loss coefficient is about 0.2 dB/km
and dispersion is about 17 ps/(km $\cdot$ nm) at 1550 nm region, we
report a QKD experiment over 260 km of this standard telecom fiber
with 52.9 dB channel loss. This is the first QKD experiment
exceeding 50 dB in channel loss and 250 km in length.

\begin{figure}[htb]
\centerline{
\includegraphics[width=8.3cm]{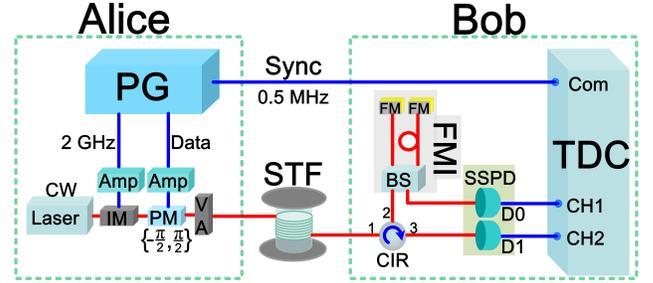}}
\caption{(Color online) Schematic of the DPS-QKD setup.} \label{exp}
\end{figure}

We chose the differential phase shift QKD (DPS-QKD) protocol
\cite{dps} to be implemented with 2 GHz rate. The experimental setup
is outlined in Fig. \ref{exp}, including the transmitter -- Alice,
quantum channel, and receiver -- Bob. At Alice's site, a continuous
wave (CW) laser, whose central wavelength is 1560.2 nm, is first
modulated into a pulse train by an intensity modulator (IM). The
phase modulator (PM) randomly encodes
$\{-\frac{\pi}{2},\frac{\pi}{2}\}$ on each pulse, and the followed
variable attenuator (VA) attenuates the average photon number per
pulse to the optimal value. Alice's pattern generator (PG) has three
outputs -- narrow pulse with 2 GHz rate to IM, pseudo-random data to
PM and 0.5 MHz sync signals to Bob. The quantum channel is the
standard telecom fiber (STF). A 3-port optical circulator (CIR) at
Bob's site is put before his 2-GHz, 1-bit delayed Faraday-Michelson
interferometer (FMI), which makes one pulse interfere with pulses
before and after it. The two outputs of the FMI are connected with a
double-channel super-conducting single photon detector (SSPD), of
which D0 channel clicks if the phase difference between two adjacent
pulses is $0$, D1 channel clicks if the phase difference is $\pi$.
Both sync signals and electrical pulses from SSPD are sent to the
time-to-digit convertor (TDC). Once TDC records a click event, Bob
and Alice can share one sifted key bit. Note that although we
transmitted the sync signal via a cable in the lab, the best way to
transmit the sync signal is over the quantum channel.

Suppose the insertion loss (IL) of Bob's detection setup and the
effective detection efficiency (EDE) of SSPD are $\alpha_{_{IL}}$
and $\eta_{_D}$, the overall transmission and detection efficiency
between Alice and Bob can be expressed as \cite{ma}
\begin{equation} \label{eta}
\eta=\eta{_{_D}} \cdot 10^{-\frac{\alpha \cdot l +
\alpha_{_{IL}}}{10}},
\end{equation}
here $\alpha$ and $l$ are the loss coefficient and the length of the
fiber respectively. Let $\mu$ denote the average photon number per
pulse set by Alice, the probability that one click event happens is
given by
\begin{equation}
p_{click}\approx p_{signal}+p_{dark},
\end{equation}
where the signal contribution is $p_{signal}=1-e^{-\mu\eta}$, the
dark count contribution is $p_{dark}=2\cdot D\cdot t_{_{W}}$, here
$D$ is the dark count rate (DCR) of Bob's detector, and $t_{_{W}}$
is the measurement time window of the system \cite{eleni}.
Considering the dead time of the detection system $t_d$, the sifted
key rate could be expressed as \cite{td}
\begin{equation} \label{siftedkey}
R_{sifted}=f \cdot p_{click} \cdot e^{-f \cdot p_{click} \cdot t_d},
\end{equation}
where $f$ is the repetition rate of transmission. If the probability
that a signal hit the wrong detector is $e_{s}$, which is the
baseline system error rate \cite{eleni}, the quantum bit error rate
(QBER) is given by
\begin{equation} \label{qber}
e=\frac{e_s \cdot p_{signal} + e_d \cdot p_{dark}}{p_{click}},
\end{equation}
where $e_d = 0.5$ means the dark count contribution is random.
Finally, the secure key rate under general individual attacks is
given by \cite{secure}
\begin{equation} \label{securekey}
R_{secure}=R_{sifted} \cdot \{\tau - f(e)\cdot H_2(e)\},
\end{equation}
where $\tau=-(1-2\mu)log_2[1-e^2-(1-6e)^2/2]$ is the compression
factor in the privacy amplification process, $f(x)$ characterizes
the efficiency of error correction algorithm, and $H_2(x)$ is the
binary Shannon entropy. In order to get a tighter security threshold
\cite{zhangq}, $f(e)$ in this paper was chosen as 1.2. Using
equations from \eqref{eta} to \eqref{securekey}, we can maximize the
secure key rate by setting optimal $\mu$ for the specific
experimental setup.

\begin{figure}[htb]
\centerline{
\includegraphics[width=8.3cm]{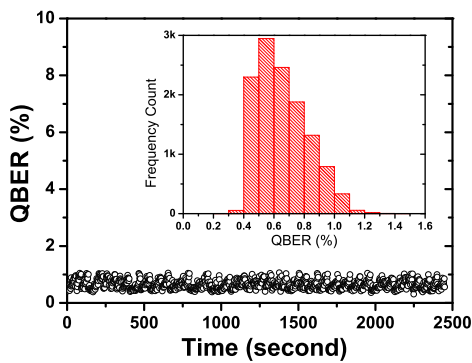}}
\caption{(Color online) Quantum bit error rate (QBER) without phase
modulation and active compensation.} \label{vis}
\end{figure}

Bob's experimental setup includes three main parts -- the
Faraday-Michelson interferometer, the super-conducting single photon
detector, and the time-to-digit convertor. \textbf{(\romannumeral
1)} One 50/50 beam splitter (BS) and two Faraday mirrors (FM)
constitute FMI, the Faraday mirror which is a combination of a 45
degree Faraday rotator and an ordinary mirror could automatically
compensate for any birefringence effect in fiber \cite{mo}, so the
2-GHz, 1-bit delayed FMI is almost insensitive to polarization.The
interferometer was insulated from the environment and actively
compensated by piezoelectric ceramics for better interference.
Without phase modulation and active compensation, the mean QBER was
about 0.65\% over 2450 seconds (Fig. \ref{vis}). However, when we
added random phase modulation signal on the phase modulator, QBER
increased to 1.80\%, which was the value of $e_{s}$ in equation
\eqref{qber}. The IL of Bob's detection setup is about 1.5 dB,
including the IL of CIR (from port 1 to port 2) and FMI.
\textbf{(\romannumeral 2)} The double-channel SSPD was made by
Scontel Ltd. from Russia, and worked with a refrigeration system
\cite{sspd} which could obtain a temperature of 1.7 K. The detector
had a counting rate more than 70 MHz, and a jitter value better than
50 ps. By carefully increasing the bias current, we achieved 3.0\%
average quantum efficiency with 1 Hz DCR. \textbf{(\romannumeral 3)}
The TDC not only recorded the sync and SSPD signals, but also set
the measurement time window $t_{_W}$ during the experiment. The
value $t_{_W}$ was set to 200 ps, and this set reduced the quantum
efficiency by 17\%. The dead time of TDC is 15 ns, which is the
value of dead time of the whole detection system.

\begin{figure}[htb]
\centerline{
\includegraphics[width=8.3cm]{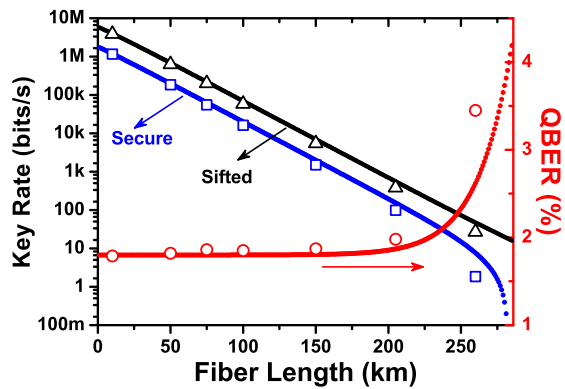}}
\caption{(Color online) Experimental results of DPS-QKD.} \label{IM}
\end{figure}

Based on these specific experimental parameters, the secure key rate
under individual attacks could be maximized by choosing optimal
$\mu$ for each fiber length, and the attainable maximum distance is
281 km (with 0.2 dB/km loss coefficient) in principle (Blue dot line
in Fig. \ref{IM}). With standard telecom fiber, the sifted key rates
and QBERs were measured at seven different fiber lengths -- 10 km,
50 km, 75 km, 100 km, 150 km, 205 km, and 260 km. We set $\mu=0.19$
for 10 km and 50 km fiber length, and $\mu=0.20$ for other length
values.

The dark count rate of detectors is a major limiting factor for
long-distance QKD \cite{cow}. Therefore the ultra low-noise SSPD was
used in our DPS-QKD system. We first set the EDE and DCR of SSPD at
2.5\% and 1 Hz respectively, and the measurement results were shown
as open shapes in Fig. \ref{IM}. For each data, we ran the QKD
process ten times, and each run lasted ten minutes. The measured
QBERs and sifted key rates are the average values of the ten runs,
but secure key rates are calculated results by equation
\eqref{securekey}. Through careful optimization of the 1-bit delayed
interferometer, we achieved the values of QBER below 2\% for the
first six lengths, and of 3.45\% for the 260 km length fiber with
52.9 dB loss (Red open circles in Fig. \ref{IM}). At 10 km and 50 km
fiber length, 1.16 Mbits/s and 185 kbits/s secure key rate under
individual attacks were achieved. At 205 km with 41.6 dB
transmission loss, 99.2 bits/s secure key rate was obtained, this
rate value was more than eight times of that achieved in 10-GHz
DPS-QKD experiment at 200 km with 42.1 dB loss \cite{nphoton}.
Although the channel loss of 260 km was one order of magnitude
larger than the loss 42.6 dB in previous 250 km QKD experiment
\cite{cow}, in which the ultra low loss fiber with 0.164 dB/km loss
coefficient was used, secure keys with 1.85 bits/s rate could still
be shared between Alice and Bob.

When the transmission fiber length was short, the signal
contribution $p_{signal}$ was much larger than the dark count
contribution $p_{dark}$. In order to get higher key rate, we
improved the quantum efficiency of SSPD by increasing the bias
current, though the dark count rate increased faster as the current
increased. In the 50 km fiber length experiment, another $\eta_{_D}
= 11.2\%$ value was tested, QBER was 1.89\%, and the corresponding
secure key rate got up to 0.81 Mbits/s, which was close to Dixon's
BB84 experiment \cite{yuan}.

For the QKD system over long distance, the nonzero accumulated
chromatic dispersion of standard telecom fiber would severely limit
the performance of QKD, especially for gigahertz systems
\cite{yuannjp}. The optical pulses are broadened during propagation
through the optical fiber, so photons would spread outside the
measurement time window, which reduces the effective quantum
efficiency, and even overlap with photons of neighbor pulses, which
degrades the encoded signal. Take the 10-GHz QKD system for example,
in which pulses are separated by 100 ps, while the dispersion is up
to 153 ps after propagating in 25 km single mode fiber \cite{eleni},
so the dispersion-shifted fiber was used in \cite{oe06} and
\cite{nphoton}. In our experiment, the full width at half maximum of
the 2-GHz pulse train was 170 ps, this wide width limited the
effects of chromatic dispersion to some extent. Fig. \ref{IM} shows
that the measured sifted key rates (or counting rates) deviate from
the simulation results, and the reduction increases as the fiber
length increases. At 205 km transmission distance, the measured
sifted key rate was 69.1\% of the simulation one. The transmission
loss 41.6 dB was higher than $0.2\times205$ dB, and the set average
photon number 0.20 was a little bit larger than the optimum one
(0.19776). After removing effects of these differences, there was
still 20.8\% reduction. The chromatic dispersion of the fiber was
the main cause of this reduction.

In summary, we have experimentally demonstrated that quantum key
distribution is possible over 260 km standard telecom fiber with
52.9 dB loss. Using the ultra low loss fiber with 0.164 dB/km loss
coefficient \cite{cow}, the quantum key exchange over 340 km
distance is in sight.

This work was supported by the National Basic Research Program of
China (Grants No. 2011CBA00200 and No. 2011CB921200), National
Natural Science Foundation of China (Grant No. 60921091), and
National High Technology Research and Development Program of China
(863 program) (Grant No. 2009AA01A349).

\end{document}